\documentclass[12pt,preprint]{aastex}

\newcommand{\hi}{H~{\sc i}}
\newcommand{\nhi}{N(H~{\sc i})}
\newcommand{\nai}{Na~{\sc i}}

\shorttitle{DIBs detected towards AO~0235+164}
\shortauthors{York et al.}

\begin{document}


\title{Detection of Diffuse Interstellar Bands in the $z=0.5$ Damped Lyman
$\alpha$ system towards AO~0235+164\footnote{Based on observations made with ESO Telescopes at the Paranal Observatories under programme ID 075.C-0029}.}


\author{Brian A. York\altaffilmark{1} and Sara L. Ellison\altaffilmark{2}}
\affil{Department of Physics and Astronomy, University of Victoria, Victoria,
BC, V8W 3P6, Canada}

\author{Brandon Lawton\altaffilmark{3} and Christopher W. Churchill\altaffilmark{4}}
\affil{Department of Astronomy, New Mexico State University, Las Cruces, NM
88003, USA}

\author{Theodore P. Snow\altaffilmark{5}}
\affil{Center for Astrophysics and Space Astronomy, University of Colorado,
Boulder, CO 80309, USA}

\author{Rachel A. Johnson\altaffilmark{6}}
\affil{Department of Astrophysics, Oxford University, Oxford, OX1 3RH, UK}

\and

\author{Sean G. Ryan\altaffilmark{7}} 
\affil{Centre for Astrophysics Research, University of Hertfordshire, 
Hatfield, AL10 9AB, UK}

\altaffiltext{1}{briany@uvic.ca}
\altaffiltext{2}{sarae@uvic.ca}
\altaffiltext{3}{blawton@nmsu.edu}
\altaffiltext{4}{cwc@nmsu.edu}
\altaffiltext{5}{Theodore.Snow@colorado.edu}
\altaffiltext{6}{raj@astro.ox.ac.uk}
\altaffiltext{7}{s.g.ryan@herts.ac.uk}


\begin{abstract}
We report the first detection of the 5705 and 5780~\AA\ Diffuse Interstellar
Bands (DIBs) in a moderate redshift Damped Lyman-$\alpha$ (DLA) system. We
measure a rest frame equivalent width of 63.2~$\pm$~8.7~m\AA\ for the 5705 and
216~$\pm$~9~m\AA\ for the 5780~\AA\ feature in the $z_{\rm abs}~=~0.524$ DLA
towards AO~0235+164 and derive limits for the equivalent widths of the bands at
5797, 6284, and 6613~\AA. The equivalent width of the 5780 band is lower than
would be expected based on the Galactic correlation of DIB strength with N(\hi),
but is in good agreement with the correlation with E(B$-$V). The relative
strengths of the 5780 and 6284~\AA\ DIBs are inconsistent with all Galactic and
extragalactic sightlines, except one Small Magellanic Cloud wing sightline
towards Sk~143. However, the relative strengths of the 5705 and 5780 \AA\ DIBs
\textit{are} consistent with the Galactic relation, indicating that the relative
strengths of these bands may be less sensitive to environment or that they may
be associated with a similar carrier.  The detection of DIBs at $z~\sim~0.5$
demonstrates that the organic compounds usually assumed to be the band carriers
were already present in the universe some five gigayears ago.
\end{abstract}




\section{INTRODUCTION}

The diffuse interstellar bands (DIBs) are a series of broad absorption lines
found between 4000 and 13000~\AA. First detected more than 80 years ago (e.g.
\citealt{Merrill:1934}), there are now several hundred known diffuse bands (e.g.
\citealt{Cox:2005}). The molecular origin (carrier) of the DIBs remains a
mystery. Amongst the potential candidates are polycyclic organic hydrocarbons
(PAHs), fullerenes, and long carbon chains \citep{Herbig:1993}.

In addition to DIB detections towards hundreds of stars in the Milky Way (see
\citealt{Herbig:1995} for a review), the DIBs have also been detected in the
Large and Small Magellanic Clouds (LMC, SMC e.g.
\citealt{Cox:2006,Ehrenfreund:2002,Welty:2006}), in NGC~1448
\citep{Sollerman:2005}, and in several starburst galaxies \citep{Heckman:2000}.
Recently, the 4428~\AA\ DIB was detected in a Damped Lyman-$\alpha$ system (a high column density \hi\ absorber with \nhi~$\ge~2.0\times~10^{20}$~cm$^{-2}$, here \nhi~=~5.0$\times$10$^{21}$cm$^{-2}$) with $z_{\rm abs}=0.524$ (DLA) towards the $z=0.94$ QSO AO~0235+164 by \citet{Junkkarinen:2004}, the first detection of a DIB at cosmological distances.

Given the Galactic correlation between \nhi\ and DIB strength (Herbig 1993),
high \nhi\ DLAs represent a promising site for DIB detection beyond the local
universe. Since DIB strength appears to be sensitive to local conditions in the
interstellar medium (ISM) such as dust-to-gas ratio and UV radiation field (e.g.
Welty et al 2006), measurements in DLAs may provide clues into the ambient
environment of high redshift galaxies. The detection of DIBs in DLAs also offers
the potential to test relationships which have been observed locally, such as
the various `families' of DIBs whose strengths show tight correlations (e.g.
\citealt{Moutou:1999}) and may be linked to the same carrier.  Observing
identical correlations in galaxies with different conditions (such as
metallicity and radiation field) is a critical test for the family hypothesis.
We have therefore embarked upon a survey for the 5780~\AA\ and other strong DIBs
in moderate redshift DLAs.	The full sample of six absorbers will be presented
in Lawton et al. (in preparation); here we present a detections of the 5705 and
5780~\AA\ feature towards one of our program QSOs, AO~0235+164. There is a
well-known, high \nhi\ DLA in this sightline at $z_{\rm abs}=0.524$ (e.g.
\citealt{Junkkarinen:2004}). This absorber is unusual in several respects: it
has a low spin temperature (\citealt{Kanekar:2003}), a relatively high
metallicity and reddening, and exhibits both the 2175~\AA\ dust feature and the
broad 4428~\AA\ DIB \citep{Junkkarinen:2004}.

\section{OBSERVATIONS AND DIB DETECTION}

We obtained six 1400-second longslit spectra of QSO AO 0235+164 using the FORS2
spectrograph at the VLT. We used the 600z grism and a 1.0 arcsecond slit which
yielded spectral coverage between 7370 and 10700~\AA, and a FWHM resolution of
5.41~\AA\ ($\sim$~3.3 pixels). At a redshift of 0.524, the spectral coverage of
the 600z grism was sufficient to search for the 5705, 5780, 5797, 6284, and
6613~\AA\ DIBs. The final S/N ratio ranged from $\sim$~60 (in the far red, which
is badly affected by night sky lines) up to $\sim$~150 per pixel. The 4428~\AA\
DIB, previously detected by \citet{Junkkarinen:2004}, was not
covered by our data.

The data were reduced using standard procedures and IRAF routines as described
in Lawton et al. (in preparation). Neither the 5705 nor the 5780~\AA\ DIB are
directly contaminated by strong sky lines within three resolution elements.
In Table~1 we give the measured rest frame equivalent widths (EWs) or 5~$\sigma$
upper limits (where we paid special attention to the impact of night sky lines,
see Lawton et al., in preparation) for the strong DIBs covered by our data.
 In addition, \nai~$\lambda \lambda$~5889,~5895 was detected with rest frame
EWs~=~793,~997~m\AA\ respectively.

\section{DISCUSSION}

\subsection{The 5780 \AA\ DIB Strength}\label{ss_5780}

\citet{Herbig:1993} has shown that, in the Milky Way, a strong relationship
exists between the EW of the 5780~\AA\ DIB and the \hi\ column density of the
line of sight. This relationship has been confirmed and improved using a much
larger and more accurate Galactic database (D. York, private communication).
In Figure~2, we show the 5780 EW of Galactic sightlines, based on data from
Herbig (1993) and the fit to a more extensive dataset from Welty et al (2006)
which includes sightlines with \nhi\ $\lesssim$ 21.6.  We also
show extragalactic sightlines (see the larger online version of Table 1) and the
5780 detection towards AO 0235+164.  Even though the scatter in the 5780--\nhi\
relationship is considerable, the 5780 EW measured towards AO 0235+164 is lower
than Galactic best fit line by six times the RMS scatter.  It is possible that
this relationship scales with metallicity, which for the DLA towards AO~0235+164
may be subsolar: $Z=(0.72~\pm~0.28)Z_{\odot}$ (Junkkarinen et al. 2004) as
measured from X-ray data.  An elemental abundance for the DLA based on UV
resonance lines such as \ion{Fe}{2} or \ion{Zn}{2} would provide an important
check of the X-ray metallicity. Welty et al. (2006) have compiled the most
complete list of DIB measurements in the LMC ($Z=0.4Z_{\odot}$) and SMC
($Z=0.2Z_{\odot}$), including sightlines with \nhi\ $\lesssim 10^{22}$
cm$^{-2}$, and find that the 5780 DIB is typically a factor of eight (LMC) and
20 (SMC) times weaker than in Galactic sightlines of the same N(\ion{H}{1}).
These deficiencies cannot be compensated for by a linear metallicity scaling.
Although the abundances of metals (as raw materials for the DIB carriers) may
have some impact on \ion{H}{1} scaling relations, it is likely that other
factors also play a role in the relations and their breadth, as discussed in the
next subsection. 

Interestingly, although the LMC and SMC points have low 5780 EWs for their \nhi\
compared to the Galactic relation, Welty et al. (2006) show that the MC
sightlines are in much better agreement with the Galactic scaling relation with
E(B$-$V).	 We reproduce this relationship in Figure~2 and include the value of
E(B$-$V)=0.23 for the DLA towards AO 0235+164 derived by Junkkarinen et al.
(2004).  As for the MC sightlines, the DLA also lies close to the Galactic
points on this graph. We remind the reader that \nhi\ is a measure of the gas
phase, while E(B$-$V) measures dust.

\subsection{Relative DIB Strengths}

Some of the DIBs are found to correlate well with one another, and these sets
are  known as `families' which may occur in the same carrier (e.g. Moutou et al.
1999). However, \citet{Wszolek:2003} note that correlation alone does not
suggest that the members of the family arise from the same carrier, suggesting
that correlated DIBs might instead arise from different carriers which tend to
occur in the same environments. Those families which are thought to arise from
the same carrier are known as `spectroscopic families', and generally include
only one strong DIB. In particular, all of the other DIBs associated with the
5780~\AA\ DIB by \citet{Wszolek:2003} (such as the features at 5776 and 5795)
are weak DIBs, with typical Galactic EWs $\lesssim$~20~m\AA, considerably below
our detection limits. Although we can not test the spectroscopic families with
the current data, the relative strengths implied by the detection limits of the
other strong DIBs covered by our spectrum have implications for the environment
of the DLA towards AO~0235+164.

\citet{Cami:1997} divide the sight-lines along which DIBs are found into four
groups: $\zeta$, $\sigma$, Orion, and circum-stellar which they argue may be
understood in terms of the strength of the local UV field. In particular,
\citet{Cami:1997} find evidence that the ionization potential of the 5797~\AA\
DIB is lower than that of the 5780~\AA\ DIB, which allows them to place the
cloud types on a continuum, with the $\zeta$-type cloud having a low UV field,
and roughly equal strengths of the 5780 and 5797~\AA\ DIBs, while the
$\sigma$-type cloud has a sufficiently high UV field that the 5797~\AA\ DIB has
started to weaken, while the strength of the 5780~\AA\ DIB is still increasing.
They hypothesize that the Orion-type cloud represents an environment with a very
high UV background, where even the carrier of the 5780~\AA\ DIB has started to
be destroyed. The $\sigma$/$\zeta$ dichotomy is thus the difference between the
cloud outer skin and the inner flesh of the cloud, with self-shielding in the
skin ($\sigma$-type) dramatically reducing the UV field inside the cloud
($\zeta$-type).  Unfortunately, our limit on the EW of the 5797~\AA\ feature is
not deep enough to distinguish between $\sigma$- and $\zeta$-types, partly
because of the proximity of this feature to a night sky line. High-resolution
spectroscopy might be able to distinguish the 5797~\AA\ DIB, if present, and
give a better chance of measuring the relative ratios.  This is technically
feasible if the QSO is observed during its bright phases when its magnitude may
reach $R~\sim~15$.

Despite the lack of a deep 5797~\AA\ EW limit, it is clear that the relative DIB
ratios in the DLA (see Table~1 and Figure~3) do not match any of the Galactic
sightlines, due to the low upper limit on the 6284~\AA\ DIB. The only sightline
known to show a similarly weak 6284~\AA\ feature is Sk~143, a sightline in the
SMC.	 Ehrenfreund et al. (2002) note that this sightline is located in the
SMC wing, which is likely more protected against UV radiation than the rest of
the SMC, and is very unusual in that it exhibits traces of the 2175~\AA\ dust
bump. The DLA towards AO~0235+164 also shows the 2175~\AA\ dust bump
(Junkkarinen et al. 2004), a rarity (although not unknown) in QSO absorbers
(\citealt{Wang:2004,York:2006}). There has been considerable debate over whether
DIB strengths correlate with bump strength, with claims both for and against
(e.g. \citealt{Herbig:1993, Desert:1995}). It is clear, however, that any
correlation has a large scatter and the presence of the 2175~\AA\ bump in both
Sk~143 and AO~0235+164 may be a red herring. For example, strong DIBs have been
found in starburst galaxies where the 2175~\AA\ feature is definitely absent
\citep{Heckman:2000}.

One interesting correlation that we can probe directly is the correlation
between the 5705 and the 5780 \AA\ DIBs presented in \citet{Thorburn:2003}. In
this case, the 5705 EW is within 10 m\AA\ of that predicted by the Galactic
relationship of \citet{Thorburn:2003}, consistent with our 1$\sigma$ EW error
plus the (small) scatter in the 5705--5780 relation. This strongly suggests that
these two DIBs either react similarly to changes in environment or that they
have closely related carriers.

The correlation of 5780~\AA\ EW with E(B$-$V) and \nhi\ indicates that, in the
future, the best targets for extragalactic DIB surveys are those towards
reddened, rather than simply the high column density, sightlines. The better
correlation of 5780 with E(B$-$V) may be because the reddening combines both
information on metallicity and the UV radiation.  Most DLAs have very low values
of E(B$-$V) (\citealt{Murphy:2004,Ellison:2005}), although \citet{Wild:2006}
have recently identified a class of absorbers detected via Ca {\sc ii}
absorption which have more significant reddening.  However, the composite of the
strongest Ca {\sc ii} absorbers was best fit with an E(B$-$V)~$\sim$~0.1, which
(following the relation between reddening and 5780 EW) would still only yield a
DIB with an equivalent width of $\sim$~13--80~m\AA.  It is therefore likely that
DIBs in QSO absorption systems will only be observed in rare cases where the
reddening is high (Lawton et al., in preparation).

\section{SUMMARY}

We have detected the 5705 and 5780~\AA\ DIBs in a DLA at $z_{\rm abs}=0.524$ in
a DLA towards AO~0235+164, and determined upper limits for the 5797, 6284, and
6613~\AA\ features.  The EW of the 5780~\AA\ feature is lower for its \nhi\ than
predicted by extrapolating the relationship in \cite{Herbig:1993}, indicating
that the relationship may depend on metallicity and/or the local radiation
field. In contrast, the 5780 EW is in good agreement with Galactic sightlines
for its E(B$-$V), as has also been observed towards Magellanic Cloud sightlines.

Unusually, the 6284~\AA\ DIB has an upper limit which is lower than the EW of
the 5780~\AA\ feature, a situation which is not generally found in the Milky
Way, or extragalactic sightlines. The one documented case of a similarly weak
6284-to-5780 ratio is towards the SMC wing sightline Sk~143. That sightline has
other characteristics in common with the DLA, including the presence of the
2175~\AA\ bump in the extinction curve, and a higher metallicity than is typical
for the SMC, possibly because of a lower radiation field.  On the other hand,
the ratio of 5705-to-5780 DIBs is in close agreement with the Galactic relation.
This may be because these DIBs respond similarly to changes in environment (e.g.
they have similar ionization potentials) and/or they have similar carriers.

Observations of H$_2$ in this DLA would be particularly interesting.  The
relative populations of various rotational levels would allow us to determine
the kinetic temperature of the gas, its density and incident flux. This goal is
not possible with current instrumentation, since the Lyman and Werner bands are
in the far-UV at $z~\sim~0.5$.	 However, the Cosmic Origins Spectrograph (COS),
scheduled to be installed during the next \textit{Hubble Space Telescope}
servicing mission, would be able to detect H$_2$ in $\sim$~15 orbits (depending
on the molecular fraction and the brightness of the QSO at the time of
observation), whilst simultaneously providing a metallicity based on UV
resonance lines.  Although future detections of DIBs towards QSO absorption
lines are likely to be challenging, due to the low incidence of highly reddened
sightlines, probing high redshifts allows us to determine the epoch at which
these organic compounds could form in the ISM.	The detection of DIBs at
$z~\sim~0.5$ shows that these particular carriers were already present some five
billion years ago.



\acknowledgments

B.~A.~Y. acknowledges the the receipt of an NSERC PGS-M award which partially funded this research.



{\it Facilities:} \facility{VLT (FORS)}.

\clearpage



\begin{figure}
\epsscale{0.6}
\plotone{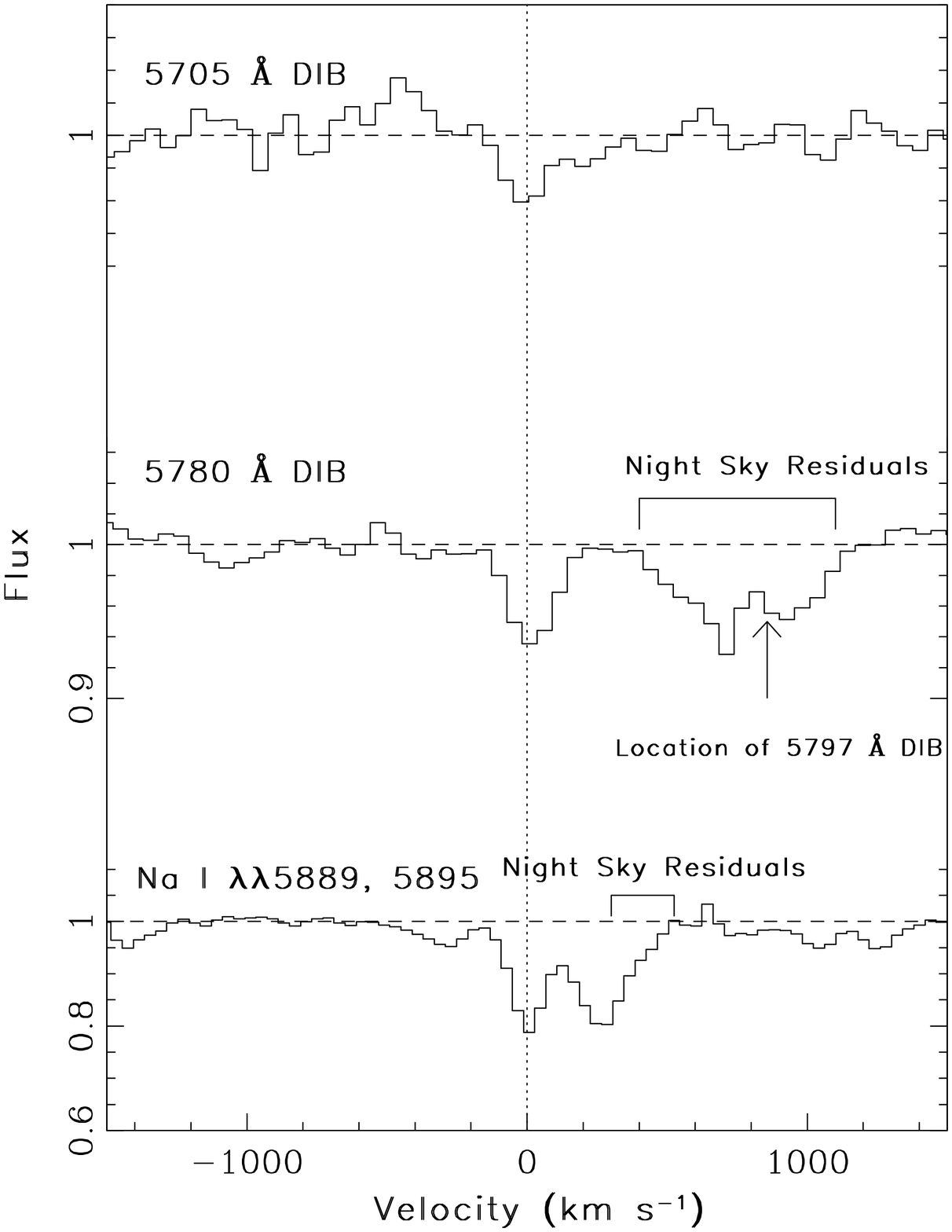}
\caption{Detection of the 5705~\AA\ DIB (upper spectrum), 5780 (middle spectrum,
with the location at which the 5797 feature would be located marked) and \nai\
doublet (lower spectrum, centred on the \nai\ $\lambda$~5889 component) in the
DLA towards AO~0235+164. The velocity scale is relative to $z_{\rm abs}=0.5238$,
which provides the best zero point for \nai\ $\lambda$~5889. The limit for the
5797~\AA\ DIB is based on simulations taking into account the effects of
contamination from sky lines.\label{fig_DATA}}
\end{figure}

\begin{figure}
\epsscale{0.8}
\plotone{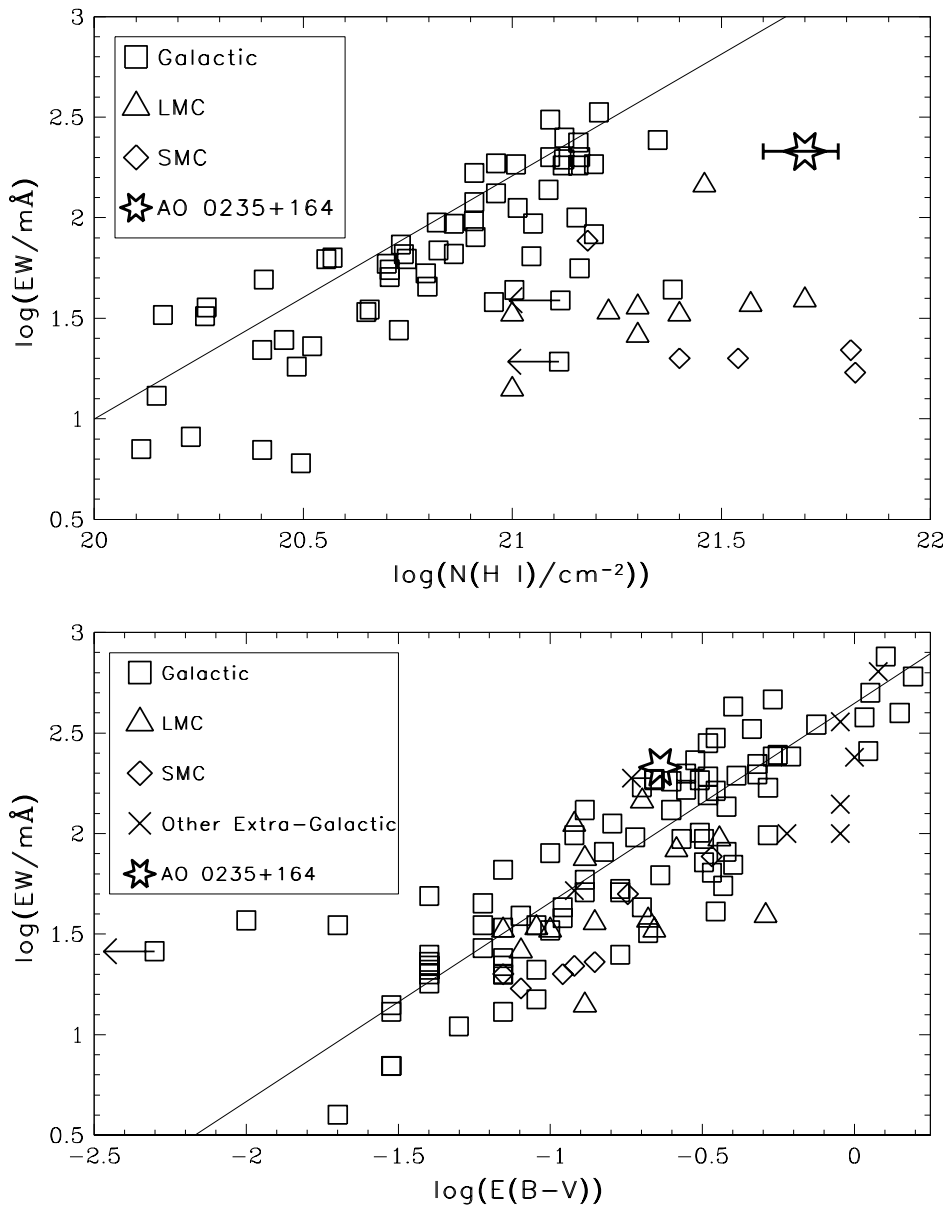}
\caption{The 5780 -- N(\hi) and the 5780 -- E(B$-$V) relationships. Galactic
points from Herbig (1993), and extragalactic points as listed in Table~1 (and
additional material online). Horizontal error bars are for a 1$\sigma$ error,
while vertical error bars are smaller than our point size. Best fit lines for
the Galactic data are from Welty et al. (2006).\label{fig_NHNA}}
\end{figure}

\begin{figure}
\epsscale{0.8}
\plotone{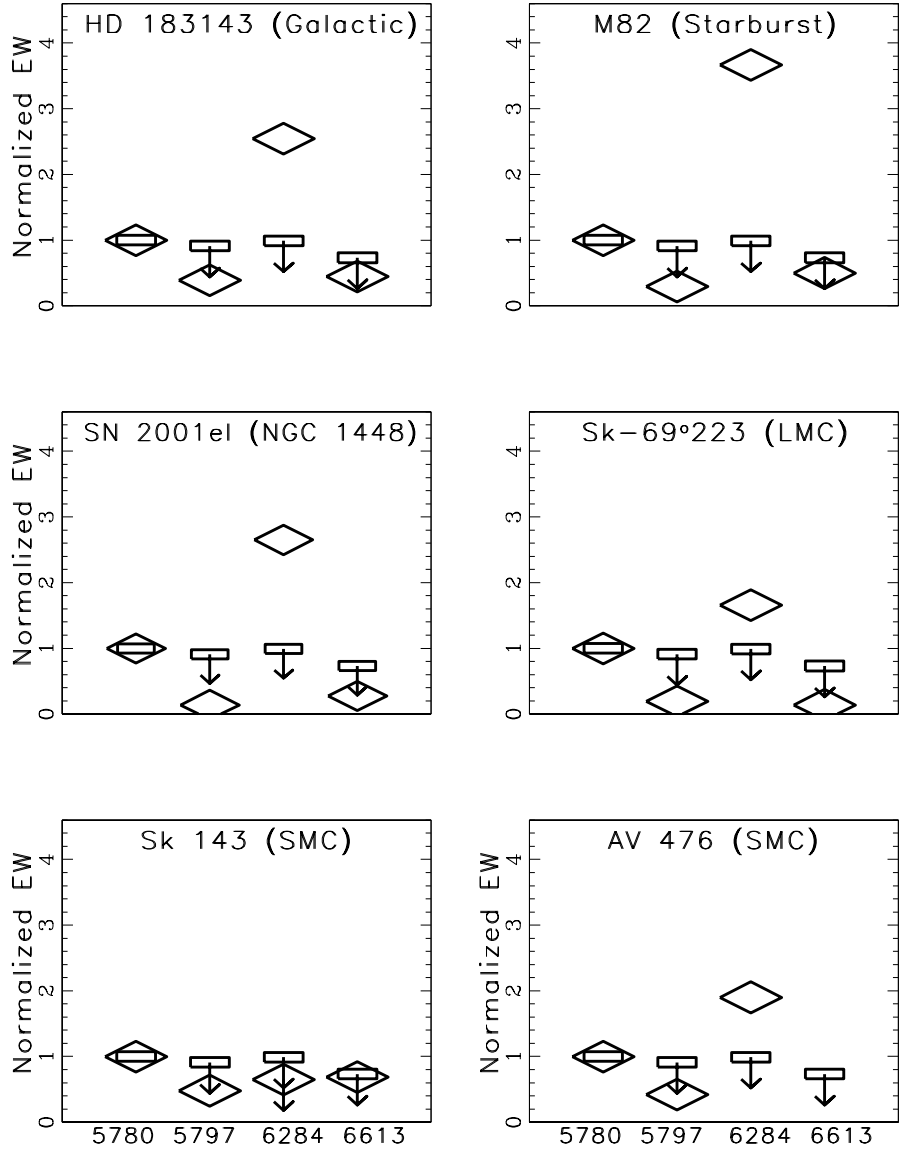}
\caption{Relative equivalent widths of the diffuse interstellar bands,
normalized to the equivalent width of the 5780~\AA\ DIB (see Table~1 for EWs).
In each case, the rectangles represent our data, while the diamonds represent
the comparison sightline.\label{fig_REL}}
\end{figure}

\clearpage

\begin{deluxetable}{ccccccccc}
\tabletypesize{\scriptsize} 
\tablecaption{\label{T_EWS}Relative Equivalent Widths of various Diffuse 
Interstellar Bands}
\tablehead{ 
\colhead{Source} &			\colhead{Location} &	\colhead{log(N(\hi))} &
\colhead{E(B$-$V)} &		\colhead{5705 \AA\ } &	\colhead{5780 \AA\ } &
\colhead{5797 \AA\ } &		\colhead{6284 \AA\ } &	\colhead{6613 \AA\ }\\
\colhead{} &				\colhead{} &			\colhead{log(cm$^{-2}$)} &
\colhead{} &				\colhead{(m\AA)} &		\colhead{(m\AA)} &		
\colhead{(m\AA)} &			\colhead{(m\AA)} &		\colhead{(m\AA)}
}
\startdata
AO 0235+164 &				DLA &					21.70$^{+0.08}_{-0.10}$ (1)&
0.23 (1) &					$63.2 \pm 8.7$ &		$216 \pm 9$ &
$< 197$ &					$< 214$ &				$< 158$ \\

Sk$-$69$^{\circ}$223 (2) &	LMC &					21.46 &
0.2 &						\nodata &				$145 \pm 2$ &
$28 \pm 6$ &				$240 \pm 21$ &			$20 \pm 8$ \\

Sk 143 (3) &				SMC &					21.18 &
0.34 &						\nodata &				$77 \pm 10$ &
$37 \pm 4$ &				$< 50$ &				$53 \pm 4$ \\

AV 476 (3) &				SMC &					\nodata &
0.18 &						\nodata &				$50 \pm 11$ &
$21 \pm 5$ &				$95 \pm 25$ &			\nodata \\

SNe 2001el (4) &			NGC 1448 &				\nodata &
0.185 &						\nodata &				189 &
26 &						500 &					52 \\

M82 (5) &					Starburst &				\nodata &
1.0 &						\nodata &				240 &
70 &						$880 \pm 30$ &			120 \\

HD 183143 (6) &				Galactic &				21.54 &
1.27 &						$172 \pm 7$ &			$758 \pm 8$ &
$295 \pm 10$ &				$1930 \pm 150$ &		$337 \pm 4$ \\
\enddata
\tablerefs{
(1) Junkkarinen et al. (2004); (2) Cox et al. (2006); (3) Welty et al. (2006);
(4) Sollerman et al. (2005); (5) Heckman and Lehnert (2000) (6) Thorburn et al.
(2003).
}
\tablecomments{All quoted limits are 5~$\sigma$ and error bars are 1~$\sigma$. A
full list of extragalactic values used in Figure~2 is given in the online
version of this table.}
\end{deluxetable}


\begin{thebibliography}{}
\bibitem[Cami et al.(1997)]{Cami:1997} Cami, J., Sonnentrucker, P.,
	Ehrenfreund, P., \& Foing, B. H., 1997, \aap, 326, 822
\bibitem[Cox et al.(2005)]{Cox:2005} Cox, N.L.J., Kaper, L., Foing, B.H.,
	\& Ehrenfreund, P., 2005, \aap, 438, 187
\bibitem[Cox et al.(2005)]{Cox:2006} Cox, N.~L.~J., Cordiner, M.~A., Cami, J.,
	Foing, B.~H., Sarre, P.~J., Kaper, L., \& Ehrenfreund, P., 2005, \aap, 
	447, 991C
\bibitem[D\'esert, Jenniskens \& Dennefeld (1995)]{Desert:1995}
		D\'esert, F.-X., Jenniskens, P., Dennefeld, M., 1995, A\&A,
		303, 223
\bibitem[Ehrenfreund et al.(2002)]{Ehrenfreund:2002} Ehrenfreund, P. et al.,
	2002, \apjl, 576, L117
\bibitem[Ellison, Hall \& Lira (2005)]{Ellison:2005}
		 Ellison, S. L., Hall, P. B., Lira, P., 2005, AJ, 130, 1345
\bibitem[Heckman \& Lehnert (2000)]{Heckman:2000} Heckman \& Lehnert, 2000,
	\apj, 537, 690
\bibitem[Herbig (1995)]{Herbig:1995} Herbig, G. H., 1995, \araa, 33, 19
\bibitem[Herbig (1993)]{Herbig:1993} Herbig, G. H., 1993, \apj, 407, 142
\bibitem[Junkkarinen et al.(2004)]{Junkkarinen:2004} Junkkarinen,ÊV.ÊT.,
	Cohen, R. D., Beaver, E. A., Burbidge, E. M., Lyons, R. W., \&
	Madejski, G., 2004, \apj, 614, 658
\bibitem[Kanekar \& Chengalur (2003)]{Kanekar:2003} Kanekar, N., \& Chengalur,
		J., 2003, A\&A, 399, 857
\bibitem[Merrill (1934)]{Merrill:1934} Merrill, P. W., 1934, PASP, 46, 206
\bibitem[Moutou et al. (1999)]{Moutou:1999} Moutou, C., Krelowski, J.,
D'Hendecourt, L.,
 Jamroszczak, J., 1999, A\&A, 351, 680
\bibitem[Murphy \& Liske (2004)]{Murphy:2004} Murphy, M. T., \& Liske, J.,
		2004, MNRAS, 354, L31
\bibitem[Sollerman et al.(2005)]{Sollerman:2005} Sollerman J., Cox, N.,
	Mattila, S., Ehrenfreund, P., Kaper, L., Leibundgut, B., \&
	Lundqvist, P., 2005, \aap, 429, 559
\bibitem[Thorburn et al.(2003)]{Thorburn:2003} Thorburn J.~A. et al., 2003, 
	\apj, 584, 339.
\bibitem[Wang et al.(2004)]{Wang:2004} Wang, J., Hall, P. B., Ge, J., \&
	Schneider, D. P., 2004, \apj, 609, 589
\bibitem[Welty et al (2006)]{Welty:2006} Welty, D. E., Federman, S. R., Gredel, 
		R., Thorburn, J. A., Lambert, D. L., 2006, ApJS accepted, 
		astro-ph/0603332	
\bibitem[Wild, Hewett \& Pettini (2006)]{Wild:2006}	 Wild, V., Hewett, P.,
		Pettini, M., 2006, MNRAS, 367, 211
\bibitem[Wszolek \& Godlowski(2003)]{Wszolek:2003} Wszolek, B., \&
	Godlowski, W., 2003, \mnras, 338, 990
\bibitem[York et al. (2006)]{York:2006}	 York, D., et al., 2006, MNRAS in press,
		astro-ph/0601279
\end{thebibliography}
\end{document}